\documentclass[letterpaper, 10 pt, conference]{ieeeconf}  % Comment this line out if you need a4paper

\IEEEoverridecommandlockouts                              % This command is only needed if 
\usepackage{graphicx}
\usepackage{amssymb}  % assumes amsmath package installed

\usepackage{xcolor} %for modification after review

\usepackage{acronym}
\usepackage{amsmath}
\usepackage{bm}

\usepackage{url}
\DeclareMathOperator*{\minimize}{\textbf{minimize}}

\newcommand{\T}[1]{\boldsymbol{\mathrm{#1}}}
\newcommand{\PCH}{\mathcal{P}}

\newacro{AD}{Autonomous Driving}
\newacro{NMPC}{Nonlinear Model Predictive Control}
\newacro{OCP}{Optimal Control Problem}
\newacro{MPC}{Model Predictive Control}
\newacro{TLS}{Truncated Legendre Series}
\newacro{MSE}{Mean Squared Error}
\newacro{LSTM}{Long-Short Term Memory}
\newacro{GRU}{Gated Recurrent Unit}
\newacro{RNNs}{Recurrent Neural Networks}
\newacro{ML}{Machine Learning}
\newacro{KPI}{Key Performance Indicator}
\newacro{KPIs}{Key Performance Indicators}
\newacro{NLP}{Nonlinear Programming Problem}
\newacro{MORRF}{MultiOutput Regressor Random Forest}
\newacro{SR}{Symbolic Regression}
\newacro{RNN}{Recurrent Neural Network}
\newacro{XAI}{eXplainable Artificial Intelligence}
\newacro{AI}{Artificial Intelligence}
\newacro{RL}{Reinforcement Learning}
\newacro{DTs}{Digital Twins}
\newacro{DT}{Digital Twin}
\newacro{CL}{closed-loop}
\newacro{OL}{open-loop}
\newacro{ExAMPC}{Explainable and Approximate NMPC}
\newacro{LUTs}{Look-Up Tables}
\newacro{CTCP}{Continuous-Time Constraint Penalties}
\newacro{AVP}{Autonomous Valet Parking}
\newacro{QP}{Quadratic Programming}
\title{\LARGE \bf
ExAMPC: the Data-Driven Explainable and Approximate NMPC with Physical Insights
}

\author{Jean Pierre Allamaa$^{1,2}$ and Panagiotis Patrinos$^{2}$ and Tong Duy Son$^{1}$%
\thanks{$^{1}$Siemens Digital Industries Software,  3001, Leuven, Belgium}%
\thanks{Email: \tt \{jean.pierre.allamaa, son.tong\}@siemens.com}%
\thanks{$^{2}$Dept. Electr. Eng. (ESAT) - STADIUS research group, KU Leuven, 3001 Leuven, Belgium}%
\thanks{Email: \tt panos.patrinos@esat.kuleuven.be}%
\thanks{This project has received funding from the European Union’s Horizon 2020 research and innovation programme under the grant agreement No. 953348 (ELO-X), and the Horizon Europe research and innovation programme under grant agreement No. 101120276 (SoliDAIR)
and the Flemish Agency for Innovation and Entrepreneurship (VLAIO) under research project No. HBC.2024.0198 (NexDT).}
\thanks{Supplementary video at: \protect\url{https://youtu.be/qGgeQaaEDXc}}
}

\begin{document}

\maketitle
\thispagestyle{empty}
\pagestyle{empty}

%%%%%%%%%%%%%%%%%%%%%%%%%%%%%%%%%%%%%%%%%%%%%%%%%%%%%%%%%%%%%%%%%%%%%%%%%%%%%%%%
\begin{abstract}
Amidst the surge in the use of Artificial Intelligence (AI) for control purposes, classical and model-based control methods maintain their popularity due to their transparency and deterministic nature. However, advanced controllers like Nonlinear Model Predictive Control (NMPC), despite proven capabilities, face adoption challenges due to their computational complexity and unpredictable closed-loop performance in complex validation systems.
This paper introduces ExAMPC, a methodology bridging classical control and explainable AI by augmenting the NMPC with data-driven insights to improve the trustworthiness and reveal the optimization solution and closed-loop performance's sensitivities to physical variables and system parameters. By employing a low-order spline embedding, we reduce the open-loop trajectory dimensionality by over 95\%, and integrate it with SHAP and Symbolic Regression from eXplainable AI (XAI) for an approximate NMPC, enabling intuitive physical insights into the NMPC's optimization routine. The prediction accuracy of the approximate NMPC is enhanced through physics-inspired continuous-time constraints penalties, reducing the predicted continuous trajectory violations by 93\%. ExAMPC also enables accurate forecasting of the NMPC's computational requirements with explainable insights on worst-case scenarios. Experimental validation on automated valet parking and autonomous racing with lap-time optimization, demonstrates the methodology's practical effectiveness for potential real-world applications.
%	
%Classical and model-based control methods maintain their popularity despite AI's rise in control applications, owing to their transparency and deterministic nature. However, advanced controllers like Nonlinear Model Predictive Control (NMPC), despite proven capabilities, face adoption challenges due to computational complexity and unpredictable closed-loop performance in complex validation systems. This paper introduces ExAMPC, a methodology bridging classical control and explainable AI by augmenting NMPC with data-driven insights to improve trustworthiness and reveal optimization solution sensitivities to physical variables and system parameters. By integrating SHAP and Symbolic Regression from eXplainable AI (XAI), and employing low-order spline embeddings, we reduce open-loop trajectory dimensionality by over 95\%, enabling intuitive physical insights into the NMPC's optimization routine. The prediction accuracy of approximate NMPC is enhanced through continuous-time constraints penalties, reducing predicted trajectory violations by 93\%. ExAMPC enables computational requirement forecasting with explainable insights into worst-case scenarios. Experimental validation on automated valet parking and autonomous racing with lap-time optimization NMPC demonstrates the methodology's practical effectiveness in real-world applications.

\end{abstract}

%%%%%%%%%%%%%%%%%%%%%%%%%%%%%%%%%%%%%%%%%%%%%%%%%%%%%%%%%%%%%%%%%%%%%%%%%%%%%%%%
\section{INTRODUCTION}
Linear \ac{MPC} stands out for its inherent explainability, allowing precise analysis of the instantaneous \ac{OL} prediction and \ac{CL} system behavior. However, this clarity on stability and performance diminishes with complex systems, such as chaotic dynamics or those involving a plant model that is more complicated than the linear prediction model in the \ac{MPC}. Moreover, \ac{MPC}'s prediction capabilities are limited by its prediction horizon, complicating the long-term \ac{CL} performance analysis trough analytical approaches.  

While data-driven control approaches such as Reinforcement Learning have gained significant traction in academia and robotics due to their minimal system knowledge requirement and their easiness to implement and use, safety-critical applications like autonomous driving demand safe, explainable, and transparent controllers. Although \ac{NMPC} inherently offers these trustworthy qualities, it poses implementation and maintenance challenges for non-experts as physical insights are required, particularly when under-performance occurs due to model mismatch or real-time operation failures. Additionally, control engineers struggle to calibrate controllers for specific \ac{CL} \ac{KPIs} that are not directly and analytically linked to the \ac{NMPC} parameters but rather emerge from the interaction of the controller with the plant, environment and other unmodelled controllers.
We propose leveraging operational \ac{CL} data to: 1) approximate the \ac{NMPC} using physics-inspired techniques, 2) expose the \ac{NMPC}'s decision making process within the specific environment conditions, and 3) predict and explain the complex system-level \ac{CL} performance \ac{KPIs}.

Training data-hungry learning-based controllers can be impractical and unsafe for systems where only operational data can be collected without disrupting the system. We propose using a small dataset to elucidate and predict system performance around current operating conditions. This approach accelerates the \ac{NMPC} design and calibration for the specific operating applications, by integrating \ac{ML}-based prediction and explainability with classical control methods like \ac{NMPC}, complementing rather than replacing \ac{AI} and \ac{NMPC} strengths. There exists several approaches allowing transparency and explainability of \ac{AI} models, known as \ac{XAI}. Those include \ac{SR}~\cite{cranmerInterpretableMachineLearning2023} that provides formulas linking outputs to input features with fast inference. Post-hoc \ac{XAI} methods like Shapley Additive exPlanations (SHAP)~\cite{lundberg2017unifiedapproachinterpretingmodel} are also beneficial for interpreting feature contributions in model predictions, particularly for Regression Trees~\cite{lundberg2020local2global}. \textcolor{black}{By calculating the features' marginal contributions affecting the prediction outcome of the ML module, SHAP renders the module's decision making process more transparent and interpretable.}

Several research works propose the combination of \ac{NMPC} with \ac{AI}. Imitation Learning of the \ac{NMPC} \ac{OL} trajectories through B-spline based coefficients embedding to penalize linear continuous-time constraints violations has been discussed by~\cite{Acerbo2020SafeIL}. 
The approximation of \ac{MPC} by relying on physics-informed constraints has been explored by~\cite{xu2024constraintsinformedneurallaguerreapproximationnonlinear}. Furthermore, Transformed-based \ac{MPC} works for generating \ac{OL} trajectories have been proposed in~\cite{2024CelestiniTransformerMPC, PARK2023108396} or in~\cite{zinage2024transformermpcacceleratingmodelpredictive} where the predicted trajectories are used to warm-start an \ac{NMPC} to accelerate its online computation. Additionally, SHAP has been applied to \ac{MPC} for model interpretation in~\cite{pmlr-v242-henkel24a}. Finally, a review in~\cite{gonzalez2024neuralnetworksfastoptimisation} discusses the use of neural networks in MPC for optimization efficiency.
Current \ac{NMPC} approximation approaches face three key challenges: 1) scalability issues and high dimensionality output demands in discrete-time sequence predictions; 2) inadequate and localized explainability of the discrete sequence element's with respect to the trend of the sequence; and 3) inability to enforce \ac{CTCP}, rendering the interpolation between two discrete points of the sequence in a possibility of constraints violation. 

We present the \ac{ExAMPC} with four main contributions: 1) proactive forecasting and monitoring of \ac{NMPC}'s \ac{CL} performance within interconnected systems, for non-experts, 2) physics-inspired \ac{NMPC} approximation using a low-order encoding via Legendre-Splines embedding, providing smooth predictions with physical insights, 3) enhanced explainability for \ac{NMPC} performance and \ac{OL} predictions through coupling with \ac{XAI} tools, and 4) experimental validation in autonomous driving and racing demonstrators.

The paper is organized as follows: in Sec.~\ref{sec:related_work} we briefly present related work on data-driven and approximate \ac{NMPC}. In Sec.~\ref{sec:prelim} we provide a background on the continuous-time optimal control problem and the employed \ac{ML} regression methods in this work. In Sec.~\ref{sec:explainable_approximate_nmpc} we introduce the low-order embedding of time-series and the physical-inspired regression model that builds the approximate NMPC for predictions with minimal continuous-time constraints violation and provide an explainability study on the \ac{OL} prediction of the approximate \ac{NMPC}. We follow with Sec.~\ref{sec:explainable_monitor} where we introduce the explainable \ac{CL} performance monitor for the \ac{NMPC} and demonstrate it on an autonomous driving and racing applications before concluding in Sec.~\ref{sec:conclusion}.

\section{Related work}\label{sec:related_work}

\begin{figure*}%
	\includegraphics[width=\textwidth,height=7cm]{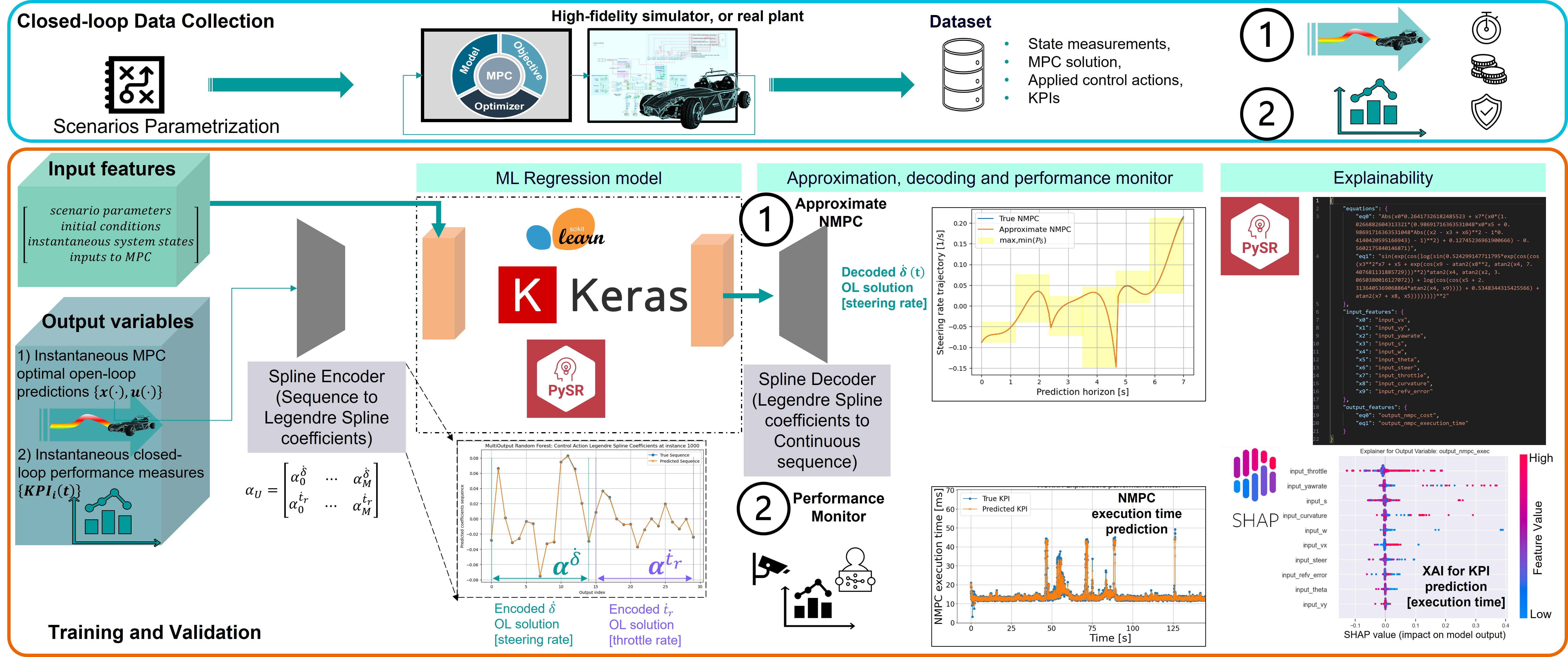}
	\caption{ExAMPC framework: employing operational closed-loop data to approximate the NMPC and a performance monitor then explain them using XAI}
	\label{fig:exampc_framework}
\end{figure*}

The importance of explainable data-driven control, as discussed in~\cite{2024RivaxDDX}, lies in its ability to enhance the transparency of the decision-making process in complex systems. Traditional approaches like Explicit \ac{MPC}~\cite{Bemporad2022ExplicitMPC}, enabled the real-time execution of the linear MPC by precomputing control laws through Multi Parametric Programming and storing them in look-up tables. However, this method remains memory-inefficient and unsuitable for nonlinear \ac{MPC}. While Explicit \ac{MPC} provides a framework for understanding system behavior in an \ac{OL} fashion and was extended to provide complexity certification for a particular set of \ac{QP} solver as in~\cite{Cimini2017ExactComplexityQP}, it lacks flexibility to systems with unknown solvers, and the adaptability to forecast performance measures such as the NMPC's computation time.
\textcolor{black} {For embedded applications, where the controller's computation time is critical, \cite{Arnstrom2024CertificationMPC} propose a method to determine the worst-case execution time of a particular class of linear MPC formulations, allowing an online monitoring and prediction of this KPI.} 
We aim at extending the previous approaches into a method allowing nonlinear constraints and dynamics handling, as well as complicated system-level \ac{CL} \ac{KPIs} \textcolor{black}{emerging from the interaction between the controller and the different system components} as in Figure~\ref{fig:exampc_framework}.

\subsection{On the use of AI for approximate NMPC}\label{subsec:ai_for_nmpc}
We focus on learning two types of output as in Figure~\ref{fig:exampc_framework}:

\begin{enumerate}
	\item the optimized \ac{NMPC} \ac{OL} trajectory for the states' and control actions' predicted evolution, which is the OCP~\eqref{eq:generic_ocp}'s solution at every instance $t$, for given state estimate $\T{x}_0(t)$:  $\{\T{x}_0(t), \eta, p(t)\} \rightarrow \{ \mathbf{x}_{(\cdot)}(t),  \mathbf{u}_{(\cdot)}(t)\}$. The autonomous \ac{CL} system evolves under uncertainties $\xi$ and scenario parameters $p(t)$ (such as target states, boundary conditions, etc..).
	\item the instantaneous measured \ac{CL} \ac{KPIs} $K_i(t)$ for $i=1,\dots,n_{KPI}$, emerging from system evolution and not necessarily analytically linked to \ac{MPC} parameters.  
\end{enumerate}
Existing approaches like~\cite{2024CelestiniTransformerMPC} that approximate the \ac{NMPC} solution to predict discrete sequences  $\T{x}_{(\cdot)}(t) = \{\T{x}(t), \T{x}(t + T_s), \dots, \T{x}(t + NT_s)\}$ over the \ac{NMPC} horizon length $N$ with a step size $T_s$, face several limitations: 
\begin{itemize}
	\item Data inefficiency:  poor scalability with large horizon length and different step sizes when using discrete methods such as direct multiple shooting, despite transformer-based technologies~\cite{2024CelestiniTransformerMPC, zinage2024transformermpcacceleratingmodelpredictive} that can handle longer trajectories, but require extensive data.
	\item High output dimensionality for discrete sequences.
	\item Lack of \ac{CTCP}, potentially compromising safety.
	\item Limited output explainability: individual sequence elements provide minimal insights into trajectory trends, dynamics, and sensitivity to input features and parameters, making them non-intuitive.
\end{itemize}
The authors of~\cite{Acerbo2020SafeIL} address the first three limitation by embedding the infinite dimensional continuous sequence $\{ \T{x}_{(\cdot)}(t),  \T{u}_{(\cdot)}(t)\}$ into a finite set of B-spline coefficients, which have a convex hull property. However, B-splines are complex to construct, highly sensitive to the chosen knot sequence, and the coefficients offer limited physical insights on the trajectory trend and dynamics. Moreover, modeling time-series with high accuracy using B-splines requires a rather dense knot sequence, where poorly chosen knots can lead to ill-conditioning in the fitting problem.

\section{Preliminaries}\label{sec:prelim}
In this section we detail the Legendre-Spline encoding based on orthogonal collocation methods, we discuss the three \ac{ML} regression methods used in this work, we set the learning objectives for approximate NMPC, and present the data generation scenarios for autonomous driving and racing.  
\subsection{Continuous-time Optimal Control Problems}
An \ac{OCP} is initially posed in continuous time and seeks to optimize a cost function $J(\T{x},\T{u})$ while satisfying a set constraints as in the nonlinear continuous Bolza problem~\cite{Huntington2007AdvancementAA}:%
\begin{align}
	\left\{
	\begin{aligned}
		\minimize_{\T{x}_{(\cdot)},\T{u}_{(\cdot)}} J(\T{x},\T{u}) &= \Phi(\T{x}(t_f)) + \int_{t_0}^{t_f} L(\T{x}(t),\T{u}(t))dt\\
		\textrm{subject to } &\dot{\T{x}}(t) = f(\T{x}(t), \T{u}(t)), \\
		&g(\T{x}(t), \T{u}(t)) \leq 0,\\
		&g_{f}(\T{x}(t_f)) \leq 0,\\
		&\T{x}(t_0) = \bar{\T{x}}_0.
	\end{aligned}
	\right.
	\raisetag{50pt}
	\label{eq:generic_ocp}
\end{align}%
where $t \in \mathbb{R}$ denotes time, $\T{x} \in \mathbb{R}^{N_x}$ is a state of the system, $\T{u} \in \mathbb{R}^{N_u}$ is a vector of control inputs. The function $\Phi : \mathbb{R}^{N_x} \rightarrow \mathbb{R}$ is the terminal cost function and $\mathit{L} : \mathbb{R}^{N_x} \times \mathbb{R}^{N_u} \rightarrow \mathbb{R}$ is the running or stage cost. The continuous-time system dynamics are given by the function $f : \mathbb{R}^{N_x} \times \mathbb{R}^{N_u} \rightarrow \mathbb{R}^{N_x}$. The function $g : \mathbb{R}^{N_x} \times \mathbb{R}^{N_u}  \rightarrow \mathbb{R}^{N_g}$ is the path linear and nonlinear constraints function, and  $g_f : \mathbb{R}^{N_x}  \rightarrow \mathbb{R}^{N_{g_f}}$ is the terminal constraints function. Finally, $\bar{\T{x}}_0$ is an input parameter to the \ac{OCP} setting the initial states condition $\T{x}(t_0)$. The \ac{OCP} in~\eqref{eq:generic_ocp} is solved in a receding horizon fashion resulting in an \ac{NMPC} \ac{CL} control framework: at every control iteration of step size $T_s$, new state measurements or estimations are fed into the \ac{OCP} to solve for~\eqref{eq:generic_ocp} over a horizon $T_H = t_f - t_0$ and the first control action $\T{u}(t_0) \rightarrow \T{u}(t_0+T_s)$ is applied. 

We define the orthogonal collocation scheme based on the \ac{TLS} of degree $M$ as in~\eqref{eq:legendre_series} to approximate the solutions $\T{x}(\tau)$ and $\T{u}(\tau)$ of the continuous time OCP in a compact representation through the coefficients $\bm{\alpha}$ rather than a discrete set of points, over the normalized time horizon $\tau\in[-1,1]$ of $t \in [t_0, t_f]$:%
\begin{equation}\label{eq:legendre_series}%
		\T{x}(\tau) = \sum_{j=0}^{M} \alpha^x_{j}\mathcal{L}_j(\tau) = (\bm{\alpha}^x)^\top\mathbf{L}_Mv(\tau).
\end{equation}%
The matrix $\mathbf{L}_M \in \mathbb{R}^{(M+1)\times(M+1)}$ is formed by the coefficients of $\mathcal{L}_j$ with respect to  $\tau$. In particular, the spanning basis are Legendre polynomials $\mathcal{L}_j(\tau)$, which have the fundamental orthogonality property. This property renders the basis terms independent from each other and less sensitive to perturbation. Moreover, the \ac{TLS} is parametrized by the coefficients $\bm{\alpha}^x = \begin{bmatrix}\alpha^x_0 & \cdots & \alpha^x_M\end{bmatrix}^\top \in \mathbb{R}^{(M+1)\times N_x}$, $\bm{\alpha}^u \in \mathbb{R}^{(M+1)\times N_u}$, and $v(\tau) = \begin{bmatrix} 1 & \tau & \tau^2 & \cdots & \tau^M \end{bmatrix}^\top$ is a vector with a geometric progression of the normalized time instance $\tau = (2t/t_f -1)$ with $t_0=0$. Equations~\eqref{eq:legendre_series} and~\eqref{eq:legendre_spline} also apply to control trajectories $\T{u}(\tau)$ with $\bm{\alpha}^u$.

For non-smooth systems and solutions that cannot be fit with a single \ac{TLS}, the normalized NMPC horizon $[-1, 1]$ can be divided into smaller finite elements to create a piecewise polynomial, namely a Legendre-Spline. On each element, every state and input is parametrized by a \ac{TLS}, the coefficients of which serve as optimization variables $\bm{\alpha}$:%
\begin{equation}%
\T{x}(\tau) = \begin{cases}
	\sum_{j=0}^{M} \alpha^x_{1, j}\mathcal{L}_j(\tau^*), & \tau_1 = -1 \leq \tau < \tau_2 \\
	\vdots \\
	\sum_{j=0}^{M} \alpha^x_{N_S, j}\mathcal{L}_j(\tau^*),  & \tau_{N_S} \leq \tau \leq 1
\end{cases}
\raisetag{0pt}
\label{eq:legendre_spline}
\end{equation}%
where $\tau^*$ converts the section limits back into $[-1, 1]$. \textcolor{black}{The normalized interval is crucial as the Legendre polynomials' definition and orthogonality properties are derived over it. }

In the generic case we refer to the Legendre-Spline coefficients of a state $x$ as $\alpha^x_{i, j}$ where $i\in[1, N_S]$ is the element number and $j\in[0, M]$ is the coefficient of order $j$ at this element. A \ac{TLS} is by design composed out of one single element, and is described by the coefficients $\alpha^x_{0, j}=\alpha^x_j$. The vector containing all the coefficients of section $i$ is denoted by $\bm{\alpha}^x_i$, and the concatenation of the coefficients over all the sections that represent the Legendre-Spline are denoted by $\bm{\alpha}^x$. Furthermore, in~\cite[Theorem 1]{ALLAMAAA2024RESAFECOL}, it is proven that the continuous-time trajectory of the state, control trajectories and linear/nonlinear constraints over them can be bounded in a finite approach through the regional convex hulls $\PCH^k$ of the Legendre-Spline through constant matrices $C_M^k$. For every \ac{TLS}, the time interval $[-1, 1]$ is divided into $K$ regions that are not necessarily equidistant, for less conservatism on the \ac{TLS}'s extrema approximation: %
\begin{subequations}%
	\label{eq:regional_safety_envelope}
	\begin{align}
	&\min\{g(\PCH_x^k, \PCH_u^k)\} - \epsilon 
	\leq g(\T{x}, \T{u})\leq 
	\max\{g(\PCH_x^k, \PCH_u^k)\} + \epsilon,\\
	&\mathrm{where} \; 	\PCH_{x}^k =  \mathcal{C}_M^k\alpha^x,\; 	\PCH_{u}^k =   \mathcal{C}_M^k\alpha^u.
	\end{align}
\end{subequations}%
%\begin{theorem}\label{thm:resafe_col}   % use the thm environment for theorems
%	Using the regional convex hull of order $M$ $\PCH_M^k$ of the state (or input) spline, the trajectory of the nonlinear constraint $g(x)$ is included in its continuous safety envelope defined by:
%	\begin{align}\label{eq:resafecol_bounded_constraint}
%		\min\{g(\PCH^k)\} - &\frac{d_k^2}{2}\max_{x\in \mathcal{X}^k}\left| \frac{d^2g}{dx^2} \right| \\ \nonumber
%		&\leq g(x)\leq \\
%		&\max\{g(\PCH^k)\} + \frac{d_k^2}{2}\max_{x\in \mathcal{X}^k}\left| \frac{d^2g}{dx^2} \right|, \nonumber
%	\end{align}
%	where $\{\mathcal{X}^k: x(t)\lvert x(t_k) \leq x(t) \leq x(t_{k+1})\}$ and $d_k = \max\{\bar{\PCH}^k_{i+1} - \bar{\PCH}^k_{i} \}$, where $\bar{\PCH}^k_{i+1}$ is a sorted $\PCH^k$ with $i=[1,\dots,M]$.
%\end{theorem}
Within the context of approximate NMPC, we propose the use of RESAFE/COL~\cite{ALLAMAAA2024RESAFECOL} to overcome the B-spline limitations presented in Sec.~\ref{subsec:ai_for_nmpc} (data efficiency, dimensionality, lack of CTCP and explainability) by using spectral collocation with orthogonal basis polynomials to achieve high solution accuracy and impose continuous-time nonlinear constraints by relying on a linear mapping between the coefficients embedding Legendre-Spline and its extrema as in~\eqref{eq:regional_safety_envelope}. This approach provides interpretable coefficients sequence revealing physical evolution information on the trajectory through the zero-order bias term $\alpha_0$ or the $i^{th}$ derivative terms. Moreover, it benefits from a better numerical conditioning and a low-order fitting: as the order $M$ increases, the high order-terms naturally vanish to zero if they do not increase the accuracy, as the coefficients $\alpha_j$ decay faster than any polynomial in $j$~\cite{boyd01Chebyshev}. We refer to this method as a low-order embedding.

\subsection{MultiOutput Regression Trees, Neural Networks and Symbolic Regression}

We aim to learn the sequence of Legendre-Spline coefficients that embed the continuous-time trajectories. The coefficients of each state trajectory are independent, but might correlate to each other between the different states (e.g. a state which is the derivative of another). For that, we employ three methods: \ac{RNN} with \ac{LSTM} layers, \ac{MORRF} and \ac{SR}. Although the focus of this paper is not about the \ac{ML} algorithm selection, we give a brief overview about the employed methods.

We implement and train \ac{RNN} using the Keras library. \ac{RNN} is chosen due to its capability to handle sequential data and capture the dependencies between the predicted sequence's elements. This is helpful as we aim to predict the NMPC's \ac{OL} trajectories, embedded as coefficients as in Figure~\ref{fig:exampc_framework}. The \ac{OL} trajectories are based on a physical system with coupling between states and controls, thus the choice of \ac{RNN}. The architecture consists of two layers (256 and 128 neurons respectively), followed by a reshaping, then an LSTM layer (64 units) and the output layer. 

The \ac{MORRF} implementation uses \textit{scikit-learn}~\cite{pedregosa2018scikitlearnmachinelearningpython}. Random Forest is an ensemble learning method that relies on the output of multiple decision trees to produce a more accurate prediction. \ac{MORRF} is suitable for small datasets that are exemplary of the current operational data and provides robust predictions with minimal hyperparameter tuning. The training of \ac{MORRF} is relatively fast and allows explainability and an almost online training then inference, which is important for engineers to rapidly understand their operating \ac{NMPC}. We use 20 estimators per Random Forest Regressor. 

While \ac{RNN} and \ac{MORRF} are non-transparent by nature, \ac{XAI} tools like~\cite{lundberg2017unifiedapproachinterpretingmodel} provide explainability to the trained algorithm, making the decision-making process of the regression model more intuitive for non-experts, on the basis that the model provides high validation and testing accuracy. This allows understanding the physical phenomena and correlations between features and outputs, which is crucial for explaining model-based controllers in \ac{CL} through \ac{XAI}.

Finally, \ac{SR} is another regression model that is explainable by design. \ac{SR} offers interpretable analytical equations linking outputs to inputs, providing physical insight to the control engineer through explicit mathematical formulas.  Moreover, the approach learns a structure of the underlying physics if the used basis functions capture the pattern well.
We use \textit{PySR}~\cite{cranmerInterpretableMachineLearning2023} with the binary operators \{$+$, $-$, $\times$, $\operatorname{atan2}(y,x)$\} and unary operators \{$\cos(x)$, $\sin(x)$, $\exp(x)$, $|x|$,  $x^2$\}, with maximum 200 iterations and 500 cycles per iteration.
We apply \ac{SR} specifically for \ac{KPIs} prediction and monitoring, where the output dimensionality is manageable.
 
\subsection{Data generation: autonomous driving and racing}
We demonstrate our work on autonomous driving and racing control applications. The NMPC is implemented in CasADi~\cite{Andersson2019Casadi} using an SQP method with OSQP as the underlying QP solver. The \ac{OCP} is transcribed into a nonlinear programming problem using RESAFE/COL~\cite{ALLAMAAA2024RESAFECOL}. The NMPC uses a fused kinematic-dynamics bicycle model with a Pacejka tire formulation. The verification uses 15 Degrees-of-Freedom high-fidelity \ac{DT} of the vehicles in Simcenter Amesim. Two demonstration setups are created:
\begin{enumerate}
	\item \ac{AVP} at speeds up to 20kph using an electric 2 seater prototype vehicle, a SimRod. The NMPC handles velocity tracking, path following and parking positioning with collision avoidance capabilities using Control Barrier Functions as in~\cite{ALLAMAAA2024RESAFECOL}. The dataset comprises 200 scenarios of 60 seconds each, featuring randomized scenario parameters for start position, parking locations, and speed. 
	\item Autonomous racing demonstrator at speeds reaching 330kph. Here the virtual \ac{NMPC} driver focuses on path tracking and lap-time optimization by maximizing the evolution along the path within a prediction horizon. The employed vehicle is a one-seater racing vehicle. A single lap around a racing track for 2 minutes and 40 seconds, sampled at 20ms, proves sufficient to demonstrate the method's effectiveness in terms of approximation and explainability with small data requirements for cases with a small operating domain.    
\end{enumerate}
Both setups integrate the \ac{NMPC} as a standalone C-code library for co-simulation with the \ac{DTs} with Simcenter Amesim for vehicle dynamics and Simcenter Prescan for environment simulation and sensor modeling for obstacle detection of crossing pedestrians and road users as visualized in Figure~\ref{fig:f1_performance_predict_1instance}. The \ac{NMPC} serves as the lowest-level control, executing trajectories from a high-level planner through steering, brake and throttle commands. 
The collected data is divided into training (64\%), validation (16\%) and testing (20\%). Finally, the data is normalized per feature and per output to $[-1, 1]$.

\section{AI as an explainable NMPC approximation}\label{sec:explainable_approximate_nmpc}
In this section we present and demonstrate the approximation of the \ac{NMPC}'s \ac{OL} solutions using \ac{ML} regression, by relying on a physics-informed, data-efficient and low-order method. Moreover, we discuss the use of \ac{XAI} techniques to gather physical insights on the optimization routine and on the trend of the \ac{OL} trajectories. Finally we demonstrate the use of \ac{ExAMPC} in the autonomous driving and racing scenarios and present the respective results.
%\begin{figure}
%	\begin{center}
%			{\includegraphics[width=0.8\columnwidth]{images/framework_for_adas}}    % The printed column width is 8.4 cm.
%			\caption{Framework for XAMPC} 
%			\label{fig:vil_on_maps} 
%		\end{center}
%\end{figure}

\subsection{RESAFE/COL for physics-informed approximate NMPC}
The NMPC solves for continuous-time trajectories that are embedded in form of Legendre-Splines' coefficients~\eqref{eq:legendre_spline}.
As explained in Sec.~\ref{subsec:ai_for_nmpc}, the use of a Legendre-Spline with orthogonal basis offers two key advantages: 1) naturally regularized low-order embedding through decoupled and independent coefficients, and 2) physical constraint enforcement via a linear coefficient mapping, enabling \ac{CTCP} without discrete sampling of the time-series trajectory.

As illustrated in Figure~\ref{fig:exampc_framework}, we collect \ac{CL} data of the NMPC controlling a high-fidelity \ac{DT} to train \ac{ML} regression models to imitate the \ac{NMPC}'s \ac{OL} solution. Using similar input parameters being fed into the \ac{NMPC}, the trained regression model would mimic the optimization process. Unlike standard imitation learning approaches that learn only the \ac{NMPC}'s policy or first control action $\T{u}(t=t_0)$, we propose to learn the complete time-series of this policy evolution $\T{u}_{(\cdot)}(t)$, and that of the states $\T{x}_{(\cdot)}(t)$. This approach captures the optimization framework linking the predicted states and control actions. It also enables an effective warm-starting strategy for the \ac{NMPC} which is known to have benefits on the numerical efficiency and helps speeding-up the computation in methods such as SQP. A baseline approach to this \ac{OL} trajectory regression predicts the sequence of the sampled discrete trajectory points, or the embedding coefficients using a \ac{MSE} loss:
\begin{equation}%
	\mathcal{L} = \mathcal{L}_{MSE} =\sum_{i=1}^{N_{batch}} \lVert(\bar{\bm{\alpha}}_i - \bm{\alpha}_i) \rVert^2/N_{batch},
\end{equation}%     
where $\bm{\alpha}_i\in \mathbb{R}^{(1\times N_{predict})}$ contains the sequence of coefficients from all the states and control respectively at instance $i$ with $N_{predict} = (M+1)\times(N_x+N_u)$ elements, and we train over batches of size $N_{batch}$, and $\bar{\bm{\alpha}}_i$ is the predicted coefficients sequence.
We enhance this loss function with a physics-informed loss using 
the convex hulls $\PCH^k$ from~\eqref{eq:regional_safety_envelope} to penalize the continuous-time constraint violations:%
\begin{subequations}
	\label{eq:physical_informed_loss}
	\begin{align}
		\mathcal{L} &= \mathcal{L}_{MSE} + \gamma \mathcal{L}_{RESAFE},\\
		\mathcal{L}_{RESAFE} &= \sum_{i=1}^{N_{bacth}}\sum_{k=1}^{K} \max(0, g(\PCH^k_{x}, \PCH^k_{u}) - \epsilon_{tol}),
	\end{align}
\end{subequations}%
where $\epsilon_{tol}$ defines the violation threshold tolerance. Note that linear state and control constraints of the form $\underline{\T{x}}\leq \T{x}(t)\leq\overline{\T{x}}$ are expressed in the generic form of $g(\T{x}, \T{u}) \leq 0$ for conciseness. The extrema elements of the convex hulls $\PCH_{u}$ and $\PCH_{v_x}$ are shown in the shaded yellow areas in Figures~\ref{fig:exampc_framework} and~\ref{fig:seq_pred_decoding}, over the decoded control actions and velocity trajectories.

\subsection{Multistep prediction using coefficients: a data efficient and explainable approach}
The proposed Legendre-Spline embedding addresses the challenges stated in Sec.~\ref{subsec:ai_for_nmpc} by encoding physical information through the coefficients: the zero order coefficient $\alpha_0$ for the trajectory bias or mean term, the first order coefficient $\alpha_1$ for information about the rate of change, the second-order coefficient $\alpha_2$ about acceleration characteristics and the higher-order terms on additional dynamic features of the trajectory. This representation enables engineers to interpret and shape the \ac{OCP}'s \ac{OL} behavior through physically meaningful parameters that can be explained using SHAP.
Therefore, by combining~\eqref{eq:legendre_spline} with~\eqref{eq:physical_informed_loss} and \ac{XAI} tools, we allow multistep sequence prediction in one shot while requiring little data due to the low dimensionality of the prediction, and while having physical insights as shown in the framework of Figure~\ref{fig:exampc_framework}.

\subsection{ExAMPC as a warm-starter for NMPC}
\begin{figure}%
	\begin{center}
		{\includegraphics[width=0.8\columnwidth]{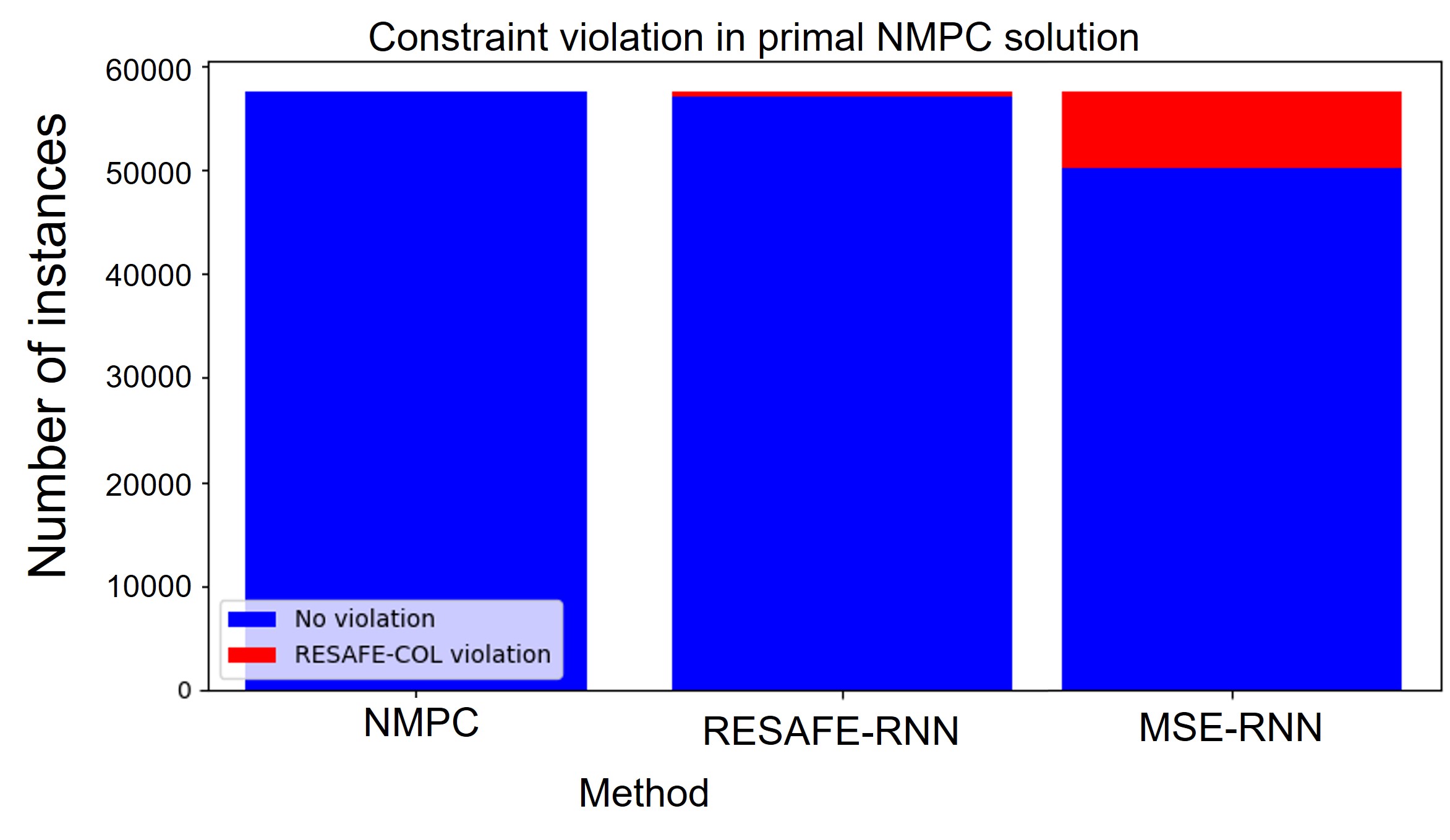}}    % The printed column width is 8.4 cm.
		\vspace{-0.2cm}
		\caption{Physics-inspired continuous-time constraints penalty with RESAFE/COL's convex hull in comparison with a baseline method} 
		\label{fig:resafecol_vs_mse} 
	\end{center}
\end{figure}

We train two \ac{RNN}s to approximate the \ac{NMPC} for the \ac{AVP} use case: one using the baseline with \ac{MSE} (MSE-RNN) and another with the RESAFE/COL-type of loss as in~\eqref{eq:physical_informed_loss} for \ac{CTCP} with $\gamma=1$ (RESAFE-RNN). As shown in Figure~\ref{fig:resafecol_vs_mse}, the approximate NMPC as RESAFE-RNN achieves 556 continuous-time constraint violations out of 57632 testing instances. The total loss is equal to 2.05e-04 divided into a \ac{MSE} of 2.0e-04 and \ac{CTCP} of 4.9e-06. 
In contrast, the baseline MSE-RNN using only coefficients learning results in 8113 violations out of 57632 instances, with a MSE on the coefficients of 1.7e-04 but a \ac{CTCP} of 8.5e-03. 
Overall, the RESAFE/COL approach with RESAFE-RNN demonstrates significant improvements equivalent to a 93\% reduction in the number of continuous-time constraint violations using the approximate NMPC. In terms of the magnitude of those violations, a reduction of 99.94\% is calculated.
\textcolor{black}{As the approximate NMPC with RESAFE-RNN 1) accurately predicts the solution of the NMPC, and 2) exhibits little constraints violations, it could be used as an initial feasible guess for the online NMPC.}
\subsection{Results and explainability for autonomous racing}
\begin{figure}%
	\begin{center}
		{\includegraphics[width=0.7\columnwidth]{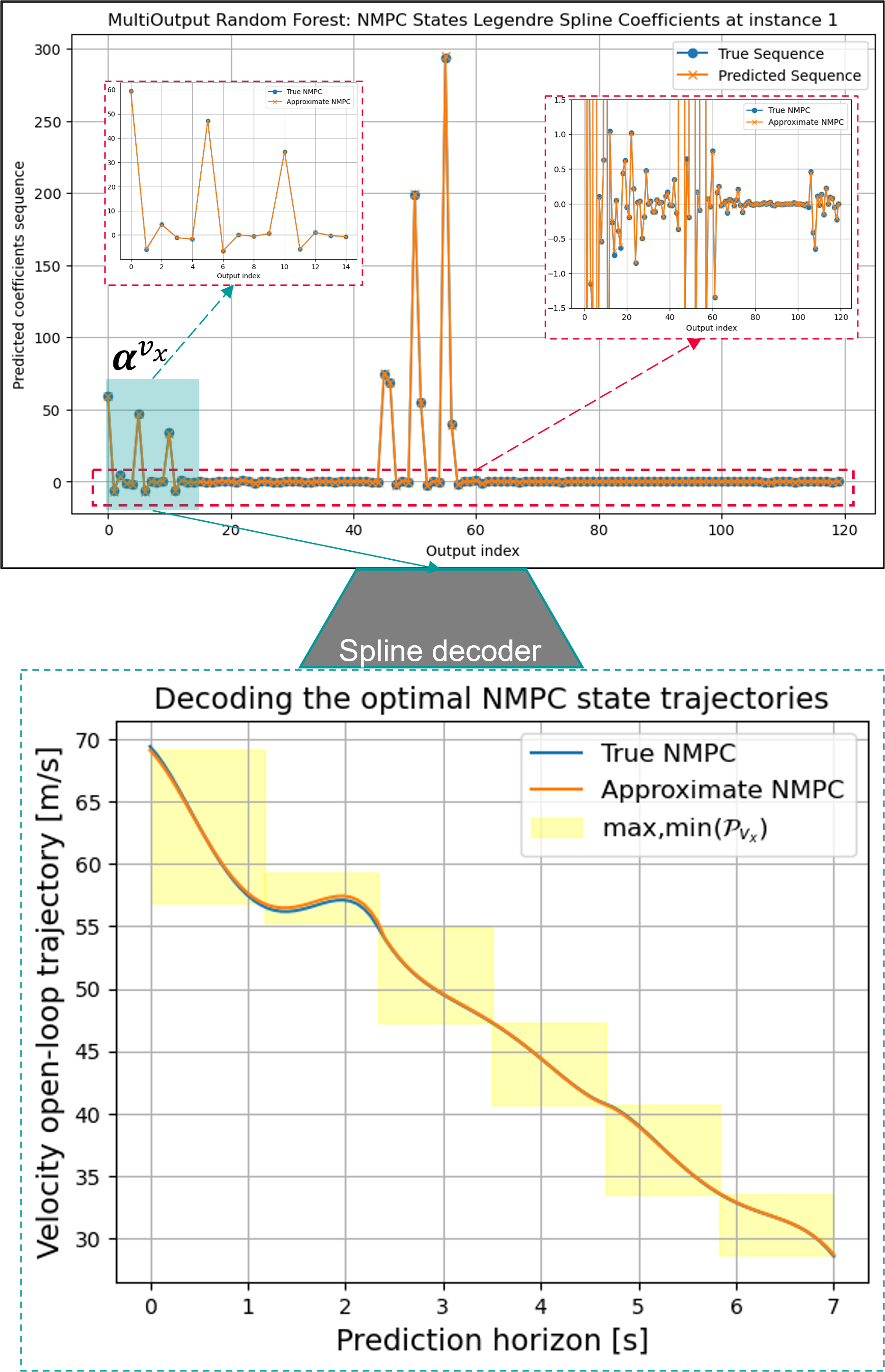}}    % The printed column width is 8.4 cm.
		\caption{Approximate NMPC: Continuous states trajectory embedding through Legendre-Spline coefficients, prediction using Random Forest Regressor, and  spline decoding with the extrema of the regional convex hulls} 
		\label{fig:seq_pred_decoding} 
	\end{center}
\end{figure}

For the autonomous racing use case, the NMPC solves for 8 states and 2 control actions \ac{OL} trajectories over $T_H = $ 7 seconds. The Legendre-Spline (c.f.~\eqref{eq:legendre_spline}, Figure~\ref{fig:seq_pred_decoding}) over the 7 seconds has $N_S=3$ sections, with an order $M=4$. That is the continuous-time trajectory of every state and control action is represented by a total of $N_{predict} =$ 15 coefficients. 
In contrast, a traditional discrete sequence prediction as in~\cite{PARK2023108396} using a sampling time $T_s=$ 20ms, would require 350 points to represent the same trajectory. The proposed method thus achieves 95.71\% reduction in dimensionality while maintaining the trajectory accuracy.

\begin{figure}%
	\begin{center}
		{\includegraphics[width=1.0\columnwidth]{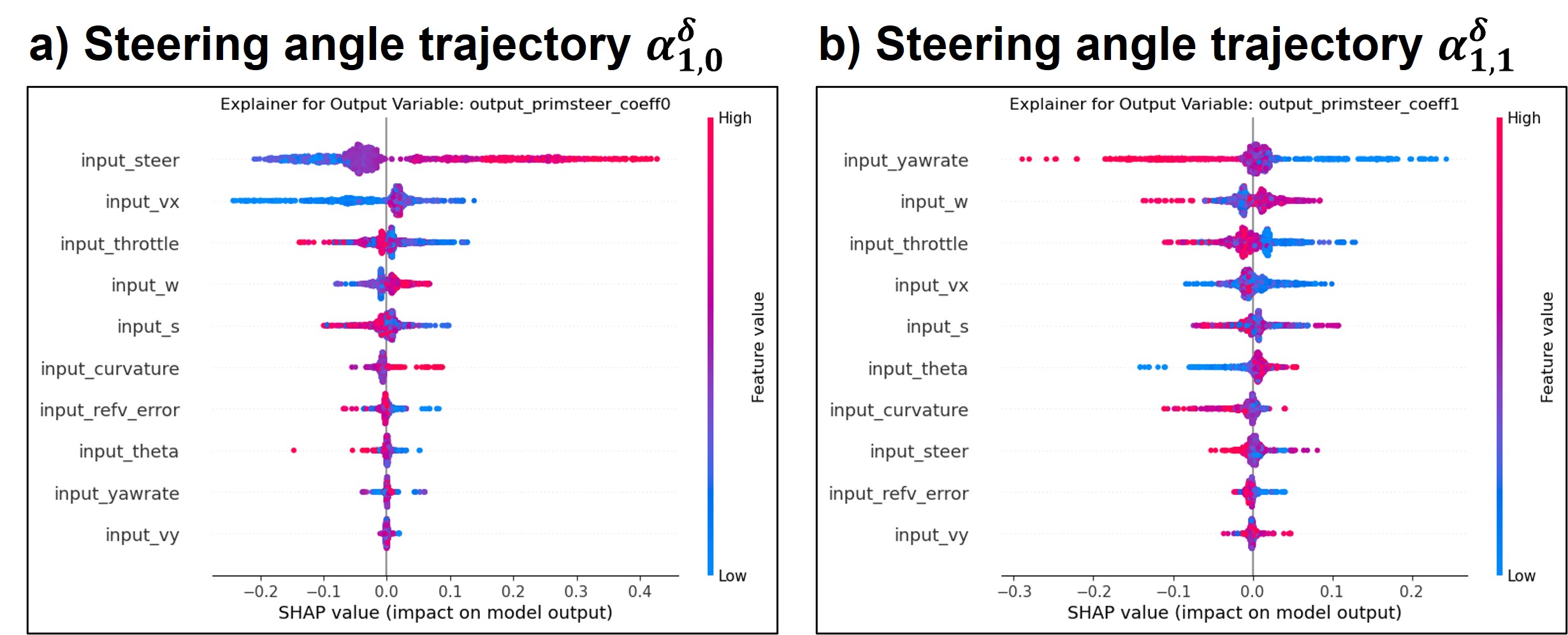}}    % The printed column width is 8.4 cm.
		\caption{Insights into the control action trends through a SHAP explainability of the approximate NMPC's Legendre-Spline coefficients} 
		\label{fig:explainable_steerthrottle_coeffs} 
	\end{center}
\end{figure}

Note that the proposed method remains compatible with discrete-time NMPC solutions such as direct multiple shooting through a least-squares fitting into Legendre-Splines. The orthogonal basis properties ensure that coefficients remain independent, leading to localized error effects. For instance, when prediction errors occur in coefficient $\alpha^{v_x}_{1,0}$, they only affect the offset of the first rolled-out \ac{TLS}, from time $t=0s$ to $t=2.33s$, while maintaining trajectory smoothness, as illustrated in Figure~\ref{fig:seq_pred_decoding}. This property stems from the orthogonal basis, where errors in $j^{th}$ coefficient only impact the corresponding $j^{th}$ derivative locally.%
The SHAP analysis reveals key insights into the \ac{OL} solution particularly for the first Legendre-Spline coefficients as illustrated in Figure~\ref{fig:explainable_steerthrottle_coeffs}, for the steering angle command $\delta$ and normalized acceleration or throttle command $t_r \in [-1,1]$ that combines the throttle and brake commands into one variable.
\textcolor{black}{SHAP values on the x-axis show each feature's impact on the model's prediction relative to the baseline (average). Positive values (right) push predictions higher, and negative values (left) lower them, with a magnitude equal to the SHAP value. Influential and important features are ordered top to bottom. Each dot represents a data point, colored, by the feature value (red for high, blue for low)}.
As the NMPC solves for the steering rate $\dot{\delta}$ and normalized throttle rate $\dot{t}_r$ as control actions $\T{u}(t)$, the current steering angle $input\_steer=\delta$ is fed as a parameter to the NMPC and approximate NMPC in the initial state estimation. For the steering trajectory, $\alpha^\delta_{1,0}$ shows the strongest explainability with the current steering measurement as in Figures~\ref{fig:explainable_steerthrottle_coeffs} (a). This is expected as the first coefficient of the sequence holds the zero-order information about the time-series which are mainly influenced by the bias term in the spline, and the OCP~\eqref{eq:generic_ocp} solves for the Legendre-Spline of the steering trajectory to be equal to $input\_steer$ at $t=0$. Notable patterns include increased steering under braking conditions (low feature value of $input\_throttle$, tending to -1). Moreover, the velocity $input\_vx$ has major impact on the output of the approximate NMPC for $\alpha^\delta_{1,0}$. The high values of steering $\alpha^\delta_{1,0}$ (both positive and negative) occur at lower velocities. This aligns with the expected behavior, as the NMPC minimizes steering at high-speeds of over 300kph to maintain path stability.
Due to orthogonality, the steering rate (Figures~\ref{fig:explainable_steerthrottle_coeffs} (b)) is mainly represented by the second coefficient of the sequence, $\alpha^\delta_{1,1}$ which carries information on the first-order derivative with respect to time. It correlates strongly with the yaw rate, showing a compensation behavior: negative yaw rates (blue or low feature value of $input\_yawrate$) trigger positive steering rates and vice versa. That is, when the vehicle is rotating counterclockwise, the NMPC reacts by steering clockwise and vice versa. Path deviations ($input\_w$) also influence steering rates, with leftward deviations (red or high feature value) triggering a clockwise (negative) corrections and steering rates.  A deceleration maneuver (negative $input\_throttle$ in blue) causes higher steering rates, mainly as the vehicle attacks corners at reduced speeds.  The predominance of left-hand corners in the racing track is reflected in the asymmetric distribution of the SHAP values in steering and steering rates towards the positive right-hand side of the plot. The orthogonality of Legendre-Spline coefficients enables this clear separation between zero-order behavior and dynamic responses, providing interpretable insights into the NMPC's decision-making process.%

%For throttle control, $\alpha^{t_r}_{1,0}$ is impacted by the current normalized throttle measurement $input\_throttle$. In fact, the OCP solves for the \ac{TLS} at the initial time $t=t_0$ to be equal to $input\_throttle$ and this is clearly illustrated in Figure~\ref{fig:explainable_steerthrottle_coeffs} (c). Moreover, $\alpha^{t_r}_{1,1}$ reveals the NMPC's lap-time optimization strategy. High throttle rates $\alpha^{t_r}_{1,1}$ (right hand of the SHAP plot in Figures~\ref{fig:explainable_steerthrottle_coeffs} (d)) occur primarily at low speeds or following braking maneuvers (low feature value of $input\_vx$ and $input\_throttle$ respectively). This indicates the controller's tendency to maximize the acceleration for minimal lap time while maintaining the vehicle stability. The orthogonality of Legendre-Spline coefficients enables this clear separation between zero-order behavior and dynamic responses, providing interpretable insights into the NMPC's decision-making process.%

\section{AI as performance monitor for the NMPC}\label{sec:explainable_monitor}
After demonstrating the approximate \ac{NMPC}, we focus in this section on the use of explainable performance monitors for system level \ac{KPIs} monitoring. The importance of employing \ac{XAI} tools such as SHAP and \ac{SR} are also highlighted as we extract important insights on the NMPC's \ac{CL} operational capability by relying on a small dataset.
%\begin{figure}
%	\begin{center}
%		{\includegraphics[width=0.8\columnwidth]{images/performance_monitor_v3}}    % The printed column width is 8.4 cm.
%		\caption{Performance monitor for the NMPC using XAI} 
%		\label{fig:performance_monitor} 
%	\end{center}
%\end{figure}
\subsection{Explainable AI for performance prediction}
An Explainable \ac{NMPC} aids users in visually understanding key parameters influencing the decision-making process, enabling real-time monitoring of the performance and suggesting when fallback controllers are necessary. As shown in Figure~\ref{fig:exampc_framework}, several instantaneous \ac{CL} \ac{KPIs} can be monitored and forecast before occurring.  We focus on two \ac{KPIs}: the NMPC optimal cost-function $K_1(t)$ indicating optimization feasibility and system energy, and the NMPC execution time $K_2(t)$ reflecting real-time computation capabilities. 
\subsection{Symbolic performance monitor}
We train \ac{MORRF} and \ac{SR} for performance prediction on $K_1, K_2$. In general, the performance prediction using \ac{MORRF} achieves superior accuracy and faster training (\ac{MSE}: 1.8e-04, quasi-instantaneous) compared to \ac{SR} (\ac{MSE}: 2.3e-03, couple of minutes). This indicates that \ac{MORRF} is able to capture better the complex coupling between those \ac{KPIs} and the input features. 
However, \ac{SR} provides explicit models linking $K_1, K_2$ to the input features by optimizing for both the structure and parameters of the model as shown in the explainability block of Figure~\ref{fig:exampc_framework}. This can enable cluster creation, and output reverse engineering. That is, if a desired computation time is to be met, an analytical operational domain of the input features can be computed by using the equations from \textit{PySR} (Figure~\ref{fig:exampc_framework}).
Furthermore, one could employ this approach to reverse engineer the designed cost function of an operating blackbox NMPC by relying on the measured features to imitate the NMPC's internal optimization. 
\subsection{Results and analysis}
\begin{figure}%
	\begin{center}
		{\includegraphics[width=1.0\columnwidth, height=2cm]{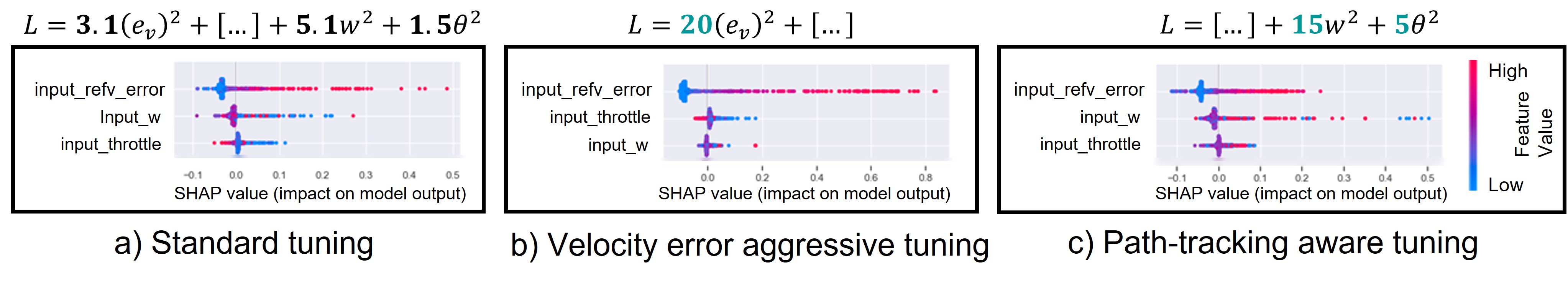}}    % The printed column width is 8.4 cm.
		\caption{Performance prediction on the cost function \ac{KPI}: reverse engineering different controller tuning in an AVP use case} 
		\label{fig:cost_monitoring_comparison_tunings} 
	\end{center}
\end{figure}

Analysis of three NMPC tuning in the \ac{AVP} demonstrator reveal interesting patterns for reverse engineering the NMPC cost function.
In Figure~\ref{fig:cost_monitoring_comparison_tunings}, the first (baseline) tuning indicates high sensitivity to velocity tracking error $input\_refv\_error$, and asymmetric responses to the path deviations $input\_w$ with deviations to the right of the path (low-feature values) impacting the cost function more than the left-hand ones (high-feature values).  Increasing the velocity tracking error weight $e_v$ from 3.1 to 20 amplifies the velocity error component's influence on the monitor prediction in the second tuning, confirming the method's ability to capture the internal NMPC optimization priorities. 
Increasing the lateral tracking error weights on path deviation $w$ and heading deviation $\theta$ in the third tuning, produces the expected quadratic cost behavior with  $input\_w$ becoming a dominant feature in the explainable monitor, where  both high and low-feature values increase the predicted NMPC cost or the consumed energy.

\begin{figure}%
	\begin{center}
		{\includegraphics[width=0.6\columnwidth]{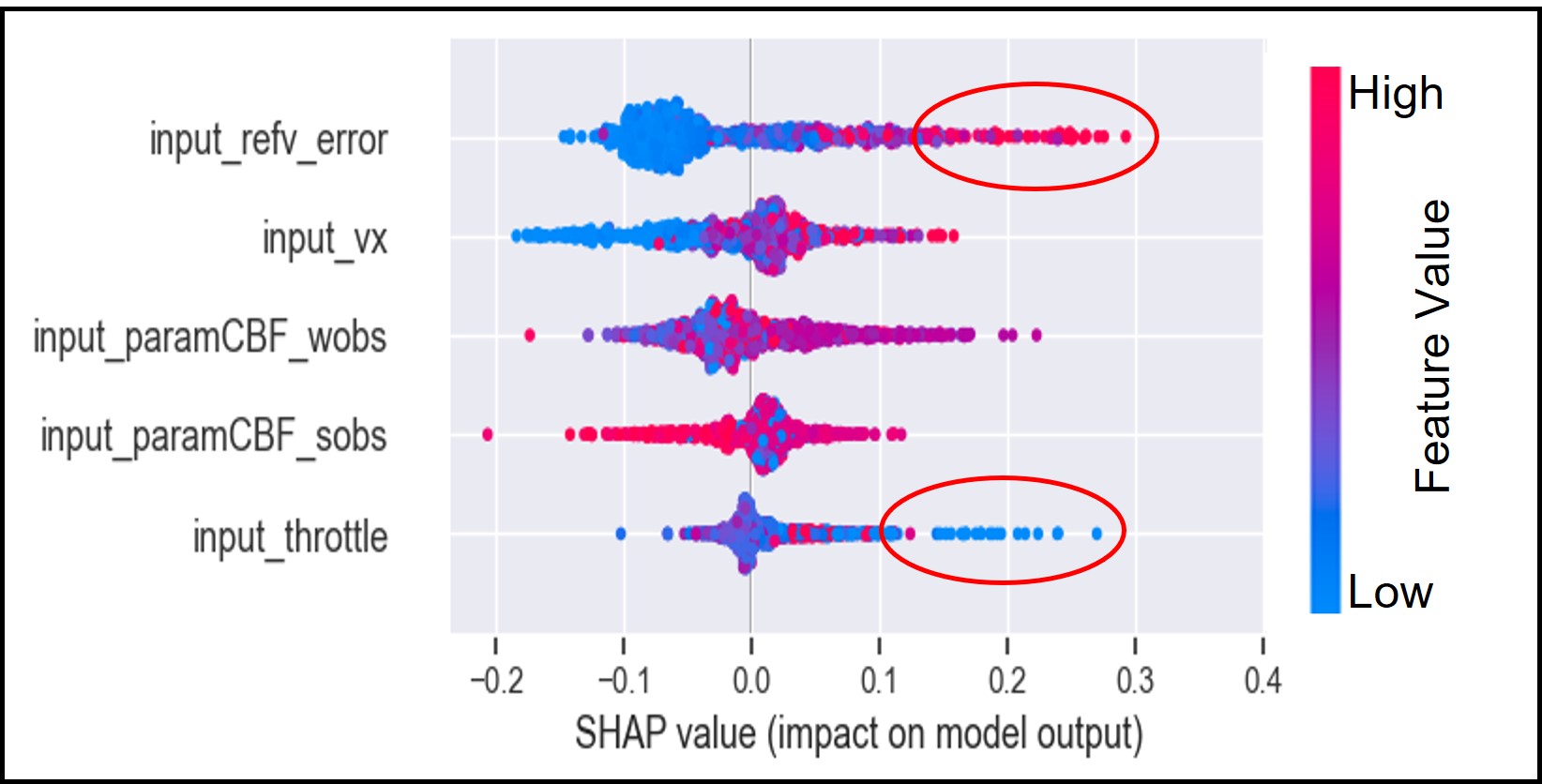}}    % The printed column width is 8.4 cm.
		\caption{Explainable performance monitor and prediction for the NMPC computation time in an AVP use case: isolating the dominant features} 
		\label{fig:avp_exec_time_predict} 
	\end{center}
\end{figure}

Explainability of the computation time prediction for the \ac{AVP} demonstrator reveals four dominant features as in Figure~\ref{fig:avp_exec_time_predict}: velocity tracking error ($input\_refv\_error$), path deviation ($input\_w$), obstacle position and heading in the frame tangent to the path ($input\_paramCBF\_*obs$), and the normalized throttle command ($input\_throttle$). Large velocity tracking errors (red) significantly increase computation time, while deceleration and braking commands (low feature value, blue) demand more computational resources than acceleration and throttling commands (high feature value, red), suggesting potential numerical challenges of the NMPC to solve at low speeds as the NMPC is more sensitive to braking ($t_r = -1$) than accelerating ($t_r=+1$).

\begin{figure}%
	\begin{center}
		{\includegraphics[width=1.0\columnwidth]{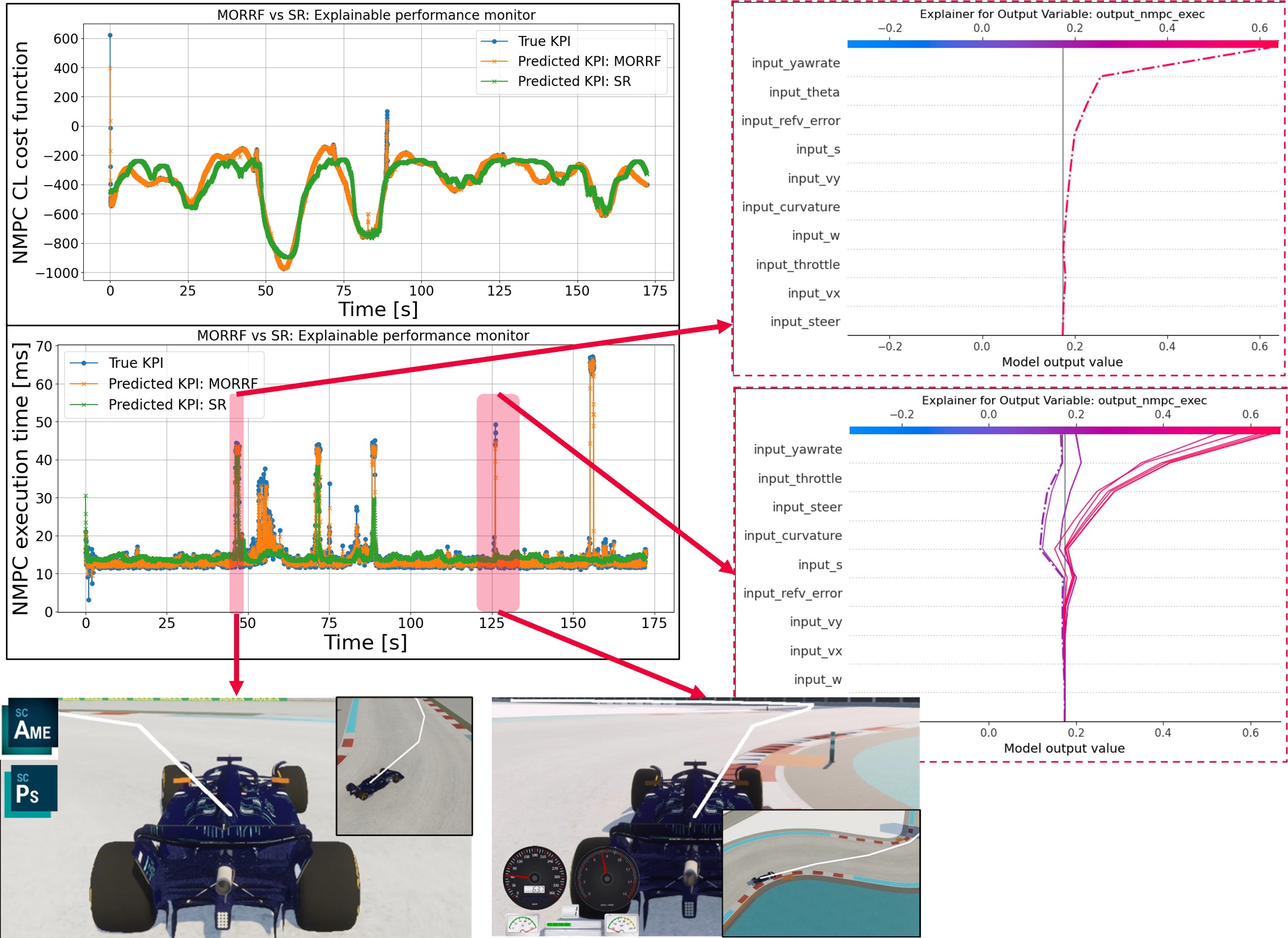}}    % The printed column width is 8.4 cm.
		\caption{Performance prediction and explainability: NMPC closed-loop optimal cost and execution time edge cases for the racing use case} 
		\label{fig:f1_performance_predict_1instance} 
	\end{center}
\end{figure}

Furthermore, we run the performance monitor for the autonomous racing demonstrator and use SHAP to understand the edge cases, as in Figure~\ref{fig:f1_performance_predict_1instance}.
\ac{MORRF} effectively captures both the \ac{NMPC}'s optimal \ac{CL} cost function value and computation time \ac{KPIs}. While the monitor accurately predicts significant cost function fluctuation as for e.g. around $t=$ 90s, the absolute computation time values are hardware-dependent and might be less generalizable. Yet, \ac{MORRF}'s capability to handle outliers, allows it to detect sudden computational peaks proving its crucial importance for the system safety monitoring. Those peaks, although rather limited, are important for edge cases studies and understanding NMPC's handling near the limits. 
A critical instance occurs near $t=$ 45s, where the vehicle exits the apex of a tight corner at 60kph before transitioning to an acceleration phase towards 330kph, as predicted by the NMPC over the next 7s. SHAP analysis of this instance reveals that yaw rate  $input\_yawrate$ and path heading deviation $input\_theta$ are the primary contributors to the well predicted high computation time. This insight provides control engineers with three actionable options: controller redesign to better handle high yaw rate scenarios, investigation of the bicycle model's numerical behavior during the optimization under large yaw rate, or implementation of a fallback controller during such challenging cornering scenarios.
Further analysis at the high-speed chicane (near $t=$ 125s) reveal more insights on the computation time patterns. While some instances maintain normal execution times around the mean, critical cases emerge from the combined effects of large normalized throttle (i.e. acceleration) and yaw rate. As shown in the \ac{DT} snapshot of Figure~\ref{fig:f1_performance_predict_1instance}, the NMPC commands a 50\% full throttle, and this occurs simultaneously under large rotation or yaw rates, as depicted in the SHAP plot. \textcolor{black}{This causes a sudden surge in computation time. The sudden activation of yaw rate and non-slip constraints in this dynamic corner scenario alters the optimal solution. This makes the warm-started primal and dual variables from the previous iteration a less effective initial guess for the new problem instance, thereby increasing the number of OSQP iterations.} Finally, the complete execution time monitor's explainability plot is presented in Figure~\ref{fig:exampc_framework}.
\section{Conclusion}\label{sec:conclusion}
This work introduces ExAMPC, an explainable and approximate NMPC and monitor framework for NMPC-controlled autonomous systems operating under model mismatch and environmental uncertainties. One aim of the work is to assist users without deep technical expertise to easily comprehend and operate an \ac{NMPC} and its behavior.
We propose the embedding of time-series through a Legendre-Spline encoding for dimensionality reduction, to approximate and explain the open-loop primal solutions of the \ac{NMPC} as state and control trajectories, through a physics-inspired loss, enhancing the continuous-time safety satisfaction by 93\%, and achieving close to no constraints violation. Additionally, by combining SHAP and Symbolic Regression, ExAMPC provides an explainable performance monitor to uncover the physical insights affecting performance indicators such as closed-loop cost and predicted energy value and the impact of measurements such as vehicle yaw rate and tracking errors on the computation time.
Future work could leverage these explainability results for targeted data generation in edge cases using \ac{DTs}, and integrate \ac{SR}-derived analytical KPI models directly into the NMPC optimization. 
%
%\addtolength{\textheight}{-12cm}   % This command serves to balance the column lengths
%                                  % on the last page of the document manually. It shortens
%                                  % the textheight of the last page by a suitable amount.
%                                  % This command does not take effect until the next page
%                                  % so it should come on the page before the last. Make
%                                  % sure that you do not shorten the textheight too much.

%%%%%%%%%%%%%%%%%%%%%%%%%%%%%%%%%%%%%%%%%%%%%%%%%%%%%%%%%%%%%%%%%%%%%%%%%%%%%%%%

%%%%%%%%%%%%%%%%%%%%%%%%%%%%%%%%%%%%%%%%%%%%%%%%%%%%%%%%%%%%%%%%%%%%%%%%%%%%%%%%

%%%%%%%%%%%%%%%%%%%%%%%%%%%%%%%%%%%%%%%%%%%%%%%%%%%%%%%%%%%%%%%%%%%%%%%%%%%%%%%%

%
\bibliographystyle{IEEEtran}
\bibliography{biblio}

\end{document}